\documentclass[sigconf]{acmart}

\AtBeginDocument{%
  }

\usepackage[draft,inline,nomargin,index]{fixme} 

\setcopyright{acmlicensed}
\acmConference[CHI '26]{CHI Conference on Human Factors in Computing Systems}
{April 13-17, 2026}{Barcelona, Spain}

\begin{document}

\title{Exploring the ``Banality'' of Deception in Generative AI}

\author{Ishitaa Narwane}
\affiliation{%
  \institution{Maastricht University}
  \city{Maastricht}
  \country{Netherlands}}
\authornote{
Corresponding author email: ishitaa.narwane@maastrichtuniversity.nl
}

\author{Johanna Gunawan}
\affiliation{%
  \institution{Maastricht University}
  \city{Maastricht}
  \country{Netherlands}}

\author{Konrad Kollnig}
\affiliation{%
  \institution{Maastricht University}
  \city{Maastricht}
  \country{Netherlands}}

\renewcommand{\shortauthors}{Narwane et al.}




\begin{abstract}
    
Current approaches to addressing deceptive design largely focus on visible interface manipulations, commonly referred to as “dark patterns.” With the rise of generative AI, deception is becoming more difficult to spot and easier to live with, as it is quietly embedded in default settings, automated suggestions, and conversational interactions rather than discrete interface elements. These subtle, normalised forms of influence, which Simone Natale \cite{natale2021deceitful} frames as ``banal deception'', shape everyday digital use and blur the line between AI-enabled assistance and manipulation.

This position paper explores \textit{banality} as a lens through which to reason through deception in generative AI experiences, especially with chatbots. We explore what \citet{natale-2025-convergence} describes as users' own involvement in their deception, and argue that this perspective could lead to future work for introducing friction to safeguard users from deception in generative AI interactions, such as empowering users through raising awareness, providing them with intervention tools, and regulatory or enforcement improvements. We present these concepts as points for discussion for the deceptive design scholarly community. 

\end{abstract}
\maketitle


\newcommand{\revdiff}[1]{{\color{black} #1}} 
\newcommand{\crdiff}[1]{{\color{black} #1}}

\renewcommand{\sectionautorefname}{\S}
\renewcommand{\subsectionautorefname}{\S}
\renewcommand{\subsubsectionautorefname}{\S}

\renewcommand{\footnotesize}{\scriptsize}
\newcommand{\etc}{etc.\xspace}
\newcommand{\eg}{e.g.,\ }
\newcommand{\etal}{et al.\xspace}
\newcommand{\ie}{i.e.,\ }
\newcommand{\re}{r.e.\ }
\newcommand{\aka}{a.k.a.\ }


\FXRegisterAuthor{in}{aix}{\color{red}Ishitaa}
\FXRegisterAuthor{jg}{aat}{\color{blue}Johanna}
\FXRegisterAuthor{kk}{ajg}{\color{orange}Konrad}



\section{Introduction and Position}
\label{Introduction}
Over the past decade, Human-Computer Interaction (HCI) scholarship and EU digital governance have focused on deceptive design tricks, often called \textit{``dark patterns''}, which lead users to make decisions that are not in accordance with their intent \cite{brignull2011dark, Gray2024}. These patterns can undermine user autonomy by leading individuals toward choices that may not align with their initial intentions through interface tricks.
This work on dark patterns has been instrumental in identifying how interface designs can trick users towards outcomes that benefit platforms over users' intentions. The harms arising from these are focused on, for example, financial loss, harms to one's privacy, or coerced consent \cite{brignull2011dark, Mathur2019}.

With the rapid adoption of generative AI (genAI) chatbots and the market dominance of GPT-based LLM chatbots developed by OpenAI, the nature of digital deception has changed significantly. Unlike many previously studied interface-based dark patterns, genAI chatbots are characterised by comparatively seamless usability (by design) that requires little to no training to start using them. That is, users only need to know how to type and how to hold a conversation to begin using genAI chatbots.
These systems are accessed through familiar and well-established digital infrastructures, such as pulling up a webpage in a browser or launching an app on their mobile device, allowing users to rely on competencies and materials they already possess rather than learning new interaction methods or acquiring specialised hardware such as VR.
Similarly, the deep integration of genAI and LLMs into already existing corporate software ecosystems helps accelerate their adoption by forcing user awareness~\cite{Rogers2003}, making chatbots nearly ubiquitous at a rate faster than other emerging technologies such as virtual or mixed reality, which have consistently encountered obstacles of their cost, hardware, and embodied interaction requirements \cite{WRZUS2024104485,Radianti2020}.

As a result, genAI chatbots move deception away from visible interface manipulation and into methods embedded in regular, everyday interactions. 
This rapid integration of ubiquitous technology poses risks ranging from short-term fraud and election tampering to long-term social-engineering attacks through the creation of realistic content and automated attack infrastructure \cite{schmitt2024digital}. This position paper uses Simone Natale~\cite{natale2021deceitful}'s concept of ``banality'' in deception as a lens to reason through deception within genAI contexts, with the aim of fostering further discussion at this workshop. We discuss this banality further in \autoref{banality-background}, expand on Natale's discussion of the role of users in banal deception in \autoref{accomplices}, and finally consider how these concepts might inform future work that empowers and protects users in \autoref{openquestions}. 
\section{The Banality of AI-Enabled Deception}
\label{banality-background}
Prior dark pattern research has established a broad ontology of dark pattern types~\cite{Gray2024} and their harms~\cite{Mathur2021} (the latter of which may include financial and privacy losses, as well as cognitive burdens, psychological distress, and violations of user agency). Some dark patterns might be more prominent, obvious, or plainly visible, whereas others are more subtle, quiet, or operate without visual cues. 
The conversational structure of genAI chatbot interactions introduces the potential for a different class of harm through conversational mechanisms: cumulative, psychological, and longitudinal effects that emerge through routine interaction rather than discrete acts of manipulation. 
We link this to \citet{natale-2025-convergence}'s concept of ``banal deception'', which articulates that deceptive mechanisms are embedded in the functioning of a technology itself. Natale also describes this embedding from the user perspective, noting that users ``actively exploit their own capacity to fall into deception''~\cite{natale2021deceitful}. %

By minimising friction and mimicking natural human conversation, these interfaces achieve the mundane, ordinary status that Natale \cite{natale-2025-convergence} considers a prerequisite for banal deception. 
The rapid evolution of these technologies implies that the technology, the user, and the interplay between both are subject to permanent change \cite{cf386dbb-f031-3590-886f-a3342f0ecf25}. The more technology disappears into the background of daily life, the more likely the user is to overlook its underlying architecture and the fact that it is still a machine even when it uses natural language \cite{natale-2025-convergence, GUZMAN2019343}. 

In this context, AI's ease of use may be seen as not just as a feature of accessibility but as the very mechanism that makes deception invisible. Banal deception can skirt user awareness and current design and legal standards due to its common yet discreet nature. 
That is, harmless appearances can, in turn, affect users' beliefs, long-term decisions, and digital trust. This mimics the historical trajectory of AI, which, since the 1950s, has explored how humans are ``programmed to be deceived'' by exploiting the limits of our perception and psychology, as Natale~\cite{natale2021deceitful} notes. 
It hides in the perceived helpfulness of a linguistic nudge or a hyper-personalised default. 
This is compounded by the generative models designed for maximum usage and friendliness. By optimising for banal objectives such as friendliness and ease of use, a medium is created in which misleading dynamics become inherent to the system's operation ~\cite{natale2021deceitful}.





\textbf{The Banality Lens in Ongoing Legal Discussions.}
The urgency of recognising banal harms in generative AI is illustrated by the ongoing \textit{Raine v OpenAI} case~\cite{adamvsopenai, RaineOpenAI2025} -- about a young teenager, Adam Raine, committing suicide. In the suit, Raine’s parents contend that ChatGPT gradually became his primary source of companionship and engaged with his suicidal ideation in ways that reinforced emotional dependency, provided detailed information related to suicide methods, creating a dependency loop that discouraged seeking real-world help, such as reaching out to family or professionals~\cite{RaineOpenAI2025}. Rather than using visible persuasive tricks, the chatbot's empathetic conversational style and engaging behaviour are designed to feel supportive and natural, which may contribute to the perceived normalisation of harmful interaction patterns. These harms could be considered banal, since chatbots are not explicitly designed to be ``evil'' but rather to be agreeable, helpful, and engaging.
Deception then occurs as the user begins to treat the AI as an emotional partner, while the AI, lacking true consciousness or ethical agency, simply reflects the user's own downward spiral to them. 
Users who become accustomed to these ongoing interactions as a result of this high level of usability may, over time, come to resemble ``prisoners of [their] own device.'' The case \textit{Raine v. OpenAI
} raises a more concerning question, that users might exploit a tool's capabilities and become active participants in their own deception ~\cite{natale-2025-convergence}. After all, children are less resilient against this capacity and in their ability to be in control of their actions.

\section{Exploring Users' ``Own Capacity'' in Deception for Future Empowerment}
\label{accomplices}
In \autoref{banality-background}, we discuss how concepts of banality argue that users are not merely passive actors but active participants in their deception \cite{natale-2025-convergence}. 
This co-production could be driven by a psychological tendency toward anthropomorphism, in which users fill in the gaps of an interface with their own social expectations. When an LLM is designed for extreme usability, it could leverage these tendencies, creating a feedback loop where the user validates the machine's human-like performance to maintain conversational flow. 

The manner in which design elements may bring users into their own deception can vary. For example, empirical work by Zhan et al.~\cite{Zhan_2025} found that LLMs exploit a truth-default state. Their study found that over-simplified responses (53.64\%) are the most frequent deceptive behaviours. By mimicking human-like cues, such as ``typing dots'' or empathetic phrasing, the AI keeps the user in a state of reflexive thinking. 

Similarly, scholars and engineers are increasingly discussing the sycophantic nature of generative models and LLMs in particular and their tendency to mirror users' agreeableness over factual accuracy to maximise human preference \cite{sharma2023towards, perez2022discoveringlanguagemodelbehaviors, wei2024simplesyntheticdatareduces,goedecke-2025-sycophancy}.

Industry standards and heuristics often describe favouring simplicity and minimising friction; Apple's design practices, for instance, are intended to be as easy-to-use as possible~\cite[p.122]{TheAppEconomy}. These are noble goals; an interface (or chatbot) that is clunky, combative, or uncomfortable to use makes for an unpleasant experience, which is generally undesirable.

In the effort to \textit{reduce} deception and resultant harms, it may seem counter-intuitive or even harsh to imply that users are complicit in their own manipulation. However, acknowledging any level of contribution -- conscious or otherwise -- to banally deceptive designs may in fact reveal opportunities for users to take back their autonomy and withdraw from this participation. If AI might be an accomplice to the proliferation of deceptive designs, and if users may also be, then we as users may be able to take back some control either through explicit awareness of banal deception or by collectively working to change this paradigm. With LLM chatbot users themselves being one half of a two-sided conversation, acknowledging their active participation may reveal more ways to mitigate the resulting harms.

The lens of banal deception articulates this paradox, in which the human-centered, highly usable social cues used to assist the user are the same cues used to deceive them. At what point does an ordinary act of assistance become an act of deception? The same questions arise when considering how the high level of usability of chatbots and generative models \textit{by design} contributes to potential deception. As such, the concept of banality may help describe why it is so difficult to draw the line between AI that assists and AI that deceives when both use the same means.
\section{Leveraging ``Banal'' Deception for User Autonomy}
\label{openquestions}
If taking banal deception as ``co-produced'' by a model's training and a user's own projection, making it difficult to assign responsibility, the concept relates to what Matthias \cite{matthias2004responsibility} describes as a ``responsibility gap'': a state where the emergent behaviour of a learning system outpaces traditional liability.
If we accept the premise that, through participation and use of a system, users participate in their own deception, how can we use this participation to strengthen user autonomy against deception?
 
Moreover, if we take the concept of co-production into account, this may change how we empower users to take back control over their use of banally deceitful LLM tools and to assert individual responsibility. Awareness, education, and community-driven intervention tools may be part of the solution, which we consider an area for future work.

Accountability research might therefore include the development of ``distributed responsibility'' models that account for the emergent, co-produced nature of AI-driven deception, along with determining how meaningful human oversight can be maintained. That is, future research acknowledging Natale's \cite{natale2021deceitful} notion of users' ``actively [exploiting] their own capacity [for] deception'' could draw on this perspective to build tools, provide education, or otherwise contribute to end-user empowerment. Our team is conducting ongoing work in the area of intervention tools for end users, as well as exploring potential mitigations from a co-production angle.

To facilitate this empowerment, future work should also support the development of enforcement tools that can detect banal deception as it occurs -- and be useful to regulators and enforcers alike in combating end users' deception. Since AI technologies evolve faster than policy, we propose a shift toward audits that monitor longitudinal interactions rather than static screenshots. Developers, cognitive and social psychologists, and policymakers must work together to design metrics that can reliably flag when a user is falling into a dependency loop as seen in \textit{Raine v. OpenAI}, before the harm becomes irreversible.

\section{Conclusion}
\label{conclusion}
This paper explores deception in generative AI through the lens of ``banality'' to discuss subtle, normalised forms of influence embedded in everyday interaction. 
Drawing on \citet{natale2021deceitful}'s work, we highlight the ordinary, mundane potential of deception and users' potential roles in their own deception, taking the position that the banality lens may contribute to dark patterns and deceptive design scholarship. 
We bring Natale's concept of banality to this workshop with the aim of encouraging discussion and reflection on the nature of AI-enabled deception.


\bibliographystyle{ACM-Reference-Format}
\bibliography{references}

\end{document}